\documentclass[letter]{aa}
\usepackage{graphicx, natbib, lscape, amssymb}
\usepackage[english]{babel}
\newcommand{\gppr}{\stackrel{>}{\scriptstyle \sim}}
\newcommand{\gappr}{\raisebox{-0.4ex}{$\gppr$}}
\newcommand{\lppr}{\stackrel{<}{\scriptstyle \sim}}
\newcommand{\lappr}{\raisebox{-0.4ex}{$\lppr$}}

\newcommand{\Porb}{\mbox{$P_\mathrm{orb}$}}

\newcommand{\Mwd}{\mbox{$M_\mathrm{wd}$}}

\newcommand{\Msun}{\mbox{$\mathrm{M}_{\odot}$}}

\begin{document}

\title{Post-common-envelope binaries from SDSS. XIII:\\
Mass dependencies of the orbital period distribution}
\titlerunning{Mass dependencies of the orbital period distribution in PCEBs}

\author{
M. Zorotovic\inst{1,2}
M.R. Schreiber\inst{1},
B.T. G\"ansicke\inst{3},
A. Rebassa-Mansergas\inst{1},
A. Nebot G{\'o}mez-Mor{\'a}n\inst{4},
J. Southworth\inst{5},
A.D. Schwope\inst{6},
S. Pyrzas\inst{3},
P. Rodr{\'i}guez-Gil\inst{7,8},
L. Schmidtobreick\inst{9},
R. Schwarz\inst{6},
C. Tappert\inst{1},
O. Toloza\inst{1},
N. Vogt\inst{1}
}
\authorrunning{M. Zorotovic et al.}
\institute{
Departamento de F\'isica y Astronom\'ia, Facultad de Ciencias, Universidad de Valpara\'iso, Valpara\'iso, Chile \\
\email{mzorotov@dfa.uv.cl}
\and
Departamento de Astronom\'ia y Astrof\'isica, Pontificia Universidad Cat\'olica de Chile, Santiago, Chile 
\and
Department of Physics, University of Warwick, Coventry, CV4 7AL, UK
\and
Universit\'e de Strasbourg, CNRS, UMR7550, Observatoire Astronomique de Strasbourg, 11 Rue de l'Universit\'e, F-67000 Strasbourg, France
\and
Astrophysics Group, Keele University, Staffordshire, ST5 5BG, UK
\and
Leibniz-Institut f\"ur Astrophysik Potsdam (AIP), An der Sternwarte 16, D-14482 Potsdam, Germany
\and
Instituto de Astrof\'isica de Canarias, V\'ia L\'actea, s/n, La Laguna, E-38205, Tenerife, Spain
\and
Departamento de Astrof\'isica, Universidad de La Laguna, E-38206 La Laguna, Tenerife, Spain
\and
European Southern Observatory, Alonso de Cordova 3107, Santiago, Chile
}
\offprints{M. Zorotovic}

\date{Received: 2 August 2011 / Accepted: 8 November 2011}

\abstract {Post-common-envelope binaries (PCEBs) consisting of a white dwarf (WD) and a main-sequence secondary star are ideal systems to constrain models of common-envelope (CE) evolution. Until very recently, observed samples of PCEBs have been too small to fully explore this potential, however the recently identified large and relatively homogenous sample of PCEBs from the Sloan Digital Sky Survey (SDSS) has significantly changed this situation.
}
{We here analyze the orbital period distributions of PCEBs containing He- and C/O-core WDs separately and investigate whether the orbital period of PCEBs is related to the masses of their stellar components.}
{We performed standard statistical tests to compare the orbital period distributions and to determine the confidence levels of possible relations.
}
{The orbital periods of PCEBs containing He-core WDs are significantly shorter than those of PCEBs containing C/O-core WDs. While the He-core PCEB orbital period distribution has a median value of $\Porb \sim 0.28$\,d, the median orbital period for PCEBs containing C/O-core WDs is $\Porb \sim 0.57$\,d. We also find that systems containing more massive secondaries have longer post-CE orbital periods, in contradiction to recent predictions. 
}
{Our observational results provide new constraints on theories of CE evolution. However we suggest future binary population models to take selection effects into account that still affect the current observed PCEB sample.
}

\maketitle

\section{Introduction}

 Since the discovery of large samples of close compact binaries \citep[]{kraft64-1}, the formation of these systems has been intensively discussed. Based on the first rough sketches provided by \citet{paczynski76-1}, the picture that most close-compact-binary stars form through common-envelope (CE) evolution has been established. Once the more massive star in the initial binary system expands on the first giant branch (FGB) or on the asymptotic giant branch (AGB), it eventually fills its Roche lobe, and dynamically unstable mass transfer to the less massive component leads to the formation of a gaseous envelope around the core of the giant and the companion. This envelope is expelled by the dissipation of orbital energy \citep[see e.g.][ for a recent review]{webbink08-1}. CE evolution terminates the mass growth of the core of the giant and therefore mass transfer initiated when the primary was on the FGB produces post-common-envelope binaries (PCEBs) containing low-mass ($\Mwd\,\lappr\,0.5$\,\Msun) He-core white dwarfs (WDs), while mass transfer starting with the primary on the AGB produces PCEBs containing more massive ($\Mwd\,\gappr\,0.5$\,\Msun) C/O-core WDs.

 Despite some recent progress, simulations of CE evolution still fail to follow the complete spiral-in and envelope ejection processes \citep[e.g.][ and references therein]{taam+ricker06-1}. However, a relation between the final and the initial orbital parameters can be obtained from a simple parametrized energy equation (e.g. \citealt{webbink84-1}; \citealt{iben+livio93-1}). A large sample of PCEBs with known stellar masses and orbital periods covering the entire parameter space, combined with any relation between the binary and/or stellar parameters of PCEBs that can be identified (e.g. a relation between the orbital period and one of the masses), might therefore provide crucial information for the energy budget of CE evolution. The potential of PCEBs to constrain current models of CE evolution was realized early by \citet{ritter86-1} and in more detail later by, e.g., \citet{schreiber+gaensicke03-1}, \citet{nelemans+tout05-1}, and \citet{davisetal10-1}. However, the PCEB samples analyzed by these authors were not homogeneous and also heavily affected by observational biases, which made any investigation of possible relations between the stellar masses and the orbital period a futile exercise.

 Over the past five years, we performed the first large-scale survey of PCEBs among white-dwarf/main-sequence (WDMS) binaries identified by the Sloan Digital Sky Survey \citep[SDSS,][]{abazajianetal09-1}. Given that the orbital period and both stellar masses can be measured relatively easily, the large and more homogeneous sample of SDSS PCEBs now available allows us for the first time to test for possible dependencies of the orbital period of PCEBs on the stellar masses. In particular, we can separate the sample according to the evolutionary state of the WD progenitor at the onset of common envelope evolution, which provides additional observational constraints for binary population models.
%This doubles the observational constraints available for binary population models.

\section{The sample}\label{sec:sample}

 The SDSS has identified large samples of WDMS binaries, and the follow-up project performed by us provides the largest and most homogeneous sample of PCEBs currently available \citep{nebot-gomez-moranetal11-1}. Potential biases affecting the SDSS PCEB sample have been analyzed with respect to the WD mass distribution \citep{rebassa-mansergasetal11-1} and the orbital period distribution \citep{nebot-gomez-moranetal11-1}. The first study concludes that there is a weak bias towards detecting PCEBs containing low-mass WDs, which does not affect the conclusions of this paper, while the second concludes that there are no significant selection effects at work related to the orbital period distribution. In contrast, both the pre-SDSS sample of PCEBs and the SDSS PCEB sample are heavily biased against systems containing early-type secondaries \citep{schreiber+gaensicke03-1,rebassa-mansergasetal10-1,schreiberetal10-1}.

 The sample of PCEBs analyzed here consists of the 62 PCEBs listed in Table\,3 of \citet{zorotovicetal11-1}, and it contains 37 new SDSS systems and 25 previously known PCEBs from the literature with accurately measured parameters. The aim of this paper is to provide observational constraints on CE evolution. In what follows we therefore use the binary parameters the systems had just after the CE was expelled, as given by \citet{zorotovicetal11-1}.

\section{Orbital period and WD core composition}\label{sec:Mwd}
\begin{figure}
\centering
\includegraphics[width=0.49\textwidth]{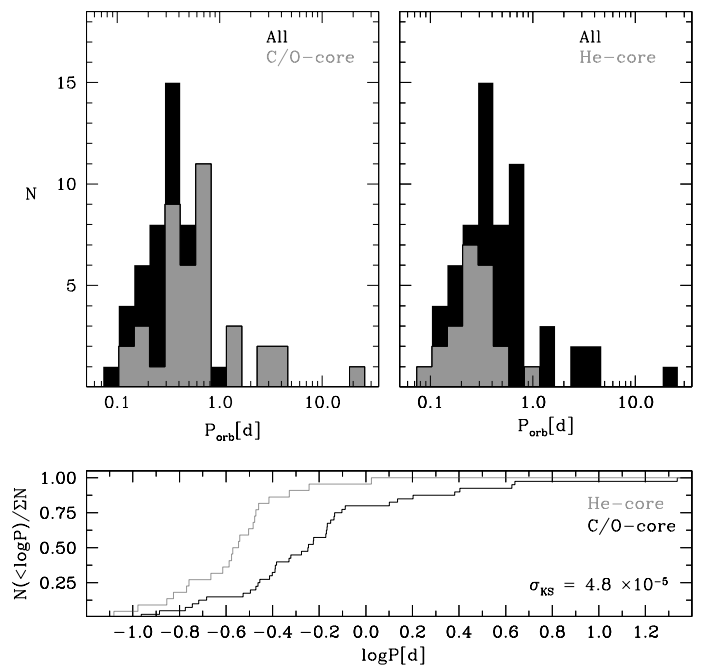}
 \caption[]{\textit{Top}: Distribution of orbital periods that the PCEBs in our sample had just after the CE phase. Systems with C/O-core WDs (\textit{left}) and He-core WDs (\textit{right}) are shown in gray, while black distributions represent the entire PCEB population. \textit{Bottom}: The cumulative distribution of post-CE periods of PCEBs containing He-core (gray) and C/O-core WDs (black). The corresponding KS probability is less than $5\times10^{-5}$, and we conclude that the two samples are not drawn from the same parent distribution.}
\label{fig:porb}
\end{figure}
 The PCEBs in our sample were separated according to the evolutionary state of the primary at the onset of CE evolution derived as in \citet{zorotovicetal11-1}. Progenitors of He-core WDs filled their Roche lobe on the FGB, while PCEBs containing C/O-core WDs descend from systems that started unstable mass transfer when the primary was on the AGB. This results in $\Mwd < 0.5$\,\Msun\, for He-core primaries and $\Mwd > 0.5$\,\Msun\, for C/O-core primaries. The top panels of Fig.\,\ref{fig:porb} show the distribution of post-CE orbital periods for the 62 PCEBs in our sample. Systems with C/O-core WDs and with He-core WDs are shown in gray in the left and right panels, respectively, while the black histograms represent the entire PCEB population. When inspecting the figure it becomes evident that systems containing He-core WDs emerged from the CE phase with systematically shorter orbital periods than PCEBs containing C/O-core WDs. The two peaks in the observed distributions have median values of $\Porb \sim 0.28$\,d and $\Porb \sim 0.57$\,d. The cumulative distributions of both samples are shown in the bottom panel and, according to a Kolmogorov-Smirnov (KS) test, are different with a confidence level of $99.995\%$.

\section{Orbital period versus WD mass}\label{sec:MwdP}
\begin{figure}
\centering
\includegraphics[width=0.49\textwidth]{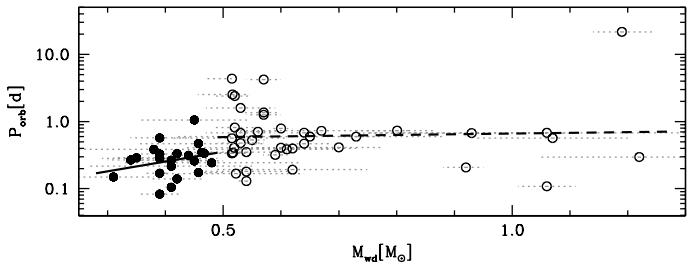}
 \caption[]{Post-CE orbital period versus WD mass for the PCEBs in our sample. Filled circles represent systems with He-core WDs and open circles systems containing CO-core WDs. The errors in the WD masses are shown as gray dotted lines. The solid and dashed lines correspond to the best linear fit for systems containing He-core and C/O-core WDs, respectively.}
\label{fig:Mwd}
\end{figure}

 Apart from the evolutionary state of the WD progenitor at the onset of mass transfer, the post-CE orbital period could also simply depend on the mass of the WD. To check this we show in Fig.\,\ref{fig:Mwd} the post-CE orbital periods of our PCEBs as a function of WD mass. Performing an F-test between a linear fit and the null hypothesis (no correlation) for systems containing He-core and C/O-core WDs\footnote{We test the two groups separately because otherwise the dependency on the core composition would affect the test. However, if there was a general trend, one would still see the same tendency in both groups.}, respectively, results in false-rejection probabilities of the null hypothesis (p-values) of $\mathrm{p}=0.35$ and $\mathrm{p}=0.79$. This implies that we cannot reject the null hypothesis; i.e., an additional term in the fit is not necessary. As a result, within neither the He-core WD nor the C/O-core WD subsample do we find statistical evidence of an increase in orbital period with primary mass. We therefore conclude that the difference between the distributions of PCEBs containing He-core and C/O-core WDs shown in Fig.\,\ref{fig:porb} is not the result of a continuous increase in the post-CE period with the mass of the WD. 

\section{Orbital period versus secondary mass}\label{sec:M2}
\begin{figure}
\centering
\includegraphics[width=0.49\textwidth]{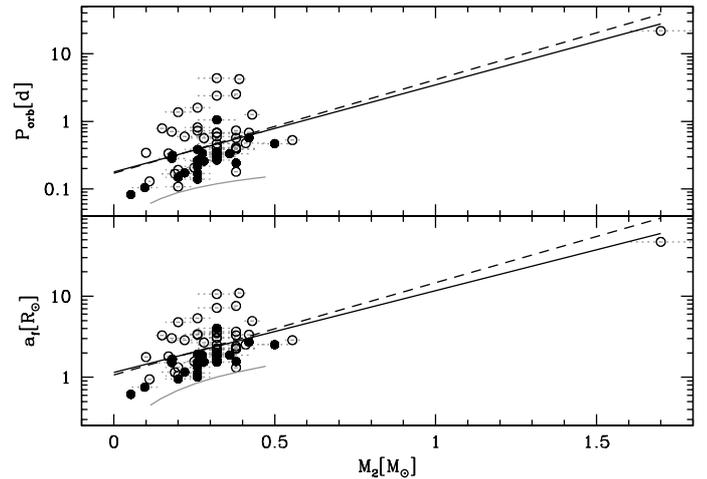}
 \caption{Post-CE orbital period (\textit{top}) and binary separation (\textit{bottom}) as a function of secondary mass for the sample of 62 PCEBs. Symbols are the same as in Fig.\,\ref{fig:Mwd}. The solid and dashed lines correspond to the best linear fit for the 62 systems and for the 60 systems with $0.08\Msun \leq M_2 \leq 1.0\Msun$ (excluding the PCEBs with the highest and lowest secondary masses), respectively, while the gray line corresponds to the CV birthline based on the mass-radius relation given in \citet{rebassa-mansergasetal07-1}.}
\label{fig:M2}
\end{figure}

 The measured stellar masses and orbital periods imply an orbital separation for each PCEB via Kepler's third law. For the relatively narrow range of WD and secondary star masses in our sample, this translates into a tight relation between orbital period and orbital separation. To test whether the orbital period (or the orbital separation) is related to the mass of the secondary, we show in Fig.\,\ref{fig:M2} the post-CE orbital period (top) and binary separation (bottom) as a function of secondary star mass. Apparently, systems containing high-mass secondaries tend to have longer orbital periods and larger separations. This trend is partly caused by the CV birthline (gray line) which, however, can not explain the absence of systems with low-mass secondaries and long periods. According to an F-test, the p-values in the top panel are $\mathrm{p}=0.03$ for all the systems and $\mathrm{p}=0.10$ in case the two extreme systems (WD$0137-3457$ and IK\,Peg) are excluded, while in the bottom panel we obtain $\mathrm{p}=0.02$ and $\mathrm{p}=0.07$, respectively. This means that there is a high probability of a real correlation, which is further supported by the only system in the sample with a massive secondary (IK\,Peg) lying very close to the linear trend identified using only PCEBs that contain less massive secondaries. However, as mentioned in Sect.\,\ref{sec:sample}, although representing a large step forward, the present PCEB sample is still heavily biased against massive companions, and the possible implications of this bias need to be carefully taken into account before drawing conclusions concerning CE evolution.

\section{Discussion}

\begin{figure}
\centering
\includegraphics[angle=270,width=0.49\textwidth]{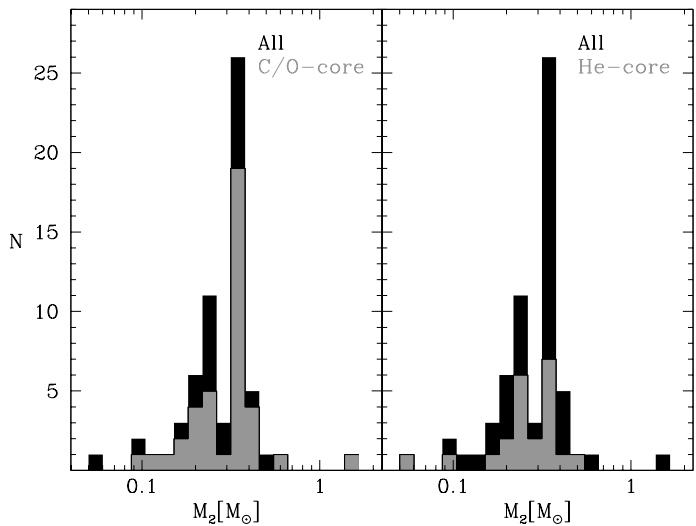}
 \caption{Distribution of secondary star masses. As in Fig.\,\ref{fig:porb} systems containing C/O-core WDs (\textit{left}) and He-core WDs (\textit{right}) are shown in gray, while black distributions represent the entire PCEB population. A $\chi^2$ test provides $Q(\chi^2)=0.59$ which implies that our data does not contain indications for the two distributions in gray to be drawn from different parent distributions.}
\label{fig:logM2}
\end{figure}

 In the previous sections we have shown that the orbital periods of PCEBs are systematically shorter for systems containing He-core WDs, are otherwise not correlated to the WD mass, and increase with the mass of the secondary star. To demonstrate that the two identified relations are independent, we show in Fig.\,\ref{fig:logM2} the distribution of secondary star masses for the whole sample of 62 PCEBs and for systems containing C/O-core or He-core WDs only (left and right panels). There seems to be no tendency toward having more massive secondaries in PCEBs with C/O-core WDs. Performing a $\chi^2$ test\footnote{We use a $\chi^2$ test instead of a KS test here because most of the systems in our sample have discrete values for the secondary mass (based on their observed spectral types). The apparent gap in the distributions at $M_2 \sim 0.3\Msun$ is not real but a consequence of the discrete values.} we find $Q(\chi^2)=0.59$, which increases to $Q(\chi^2)=0.63$ if we exclude the two extreme systems with $M_2 < 0.08\Msun$ or $M_2 > 1.0\Msun$. We therefore conclude that the significant difference between the post-CE orbital period distributions of PCEBs with He-core and C/O-core WDs (Fig.\,\ref{fig:porb}) and the increase in orbital period with secondary mass (Fig.\,\ref{fig:M2}) are independently present in our data.

\subsection{He versus C/O core}

After we exclude a continuous increase in the orbital period with WD mass and taking into account that the secondary mass distributions are
similar for systems containing He-core and C/O-core WDs, the different post-CE orbital period distributions shown in Fig.\,\ref{fig:porb} must be related to the different evolutionary stages of the primary at the onset of the unstable mass transfer that initiated the CE evolution. We see straightforward explanations for this result. First, the progenitors of C/O-core WDs started mass transfer on the AGB and must have exceeded their maximum radius on the FGB, which implies a larger initial orbital separation. Second, the binding energy parameter $\lambda$ is greater for stars on the AGB \citep[at least for low- and intermediate-mass primaries, see][]{dewi+tauris00-1}, which implies less binding energy of the envelope. This effect would be even stronger if a fraction of the internal energy, like recombination, helps to expel the envelope. This energy becomes comparable to the lower gravitational energy in more extended envelopes, and therefore its relative contribution in reducing the binding energy of the envelope (increasing $\lambda$) is greater.

\subsection{CE evolution and the secondary mass}

 Recently, \citet{demarcoetal11-1} have reconstructed the CE evolution of a sample of PCEBs and predicted a dependency of the CE efficiency $\alpha_\mathrm{CE}$ on the mass ratio, leading to a larger final separation for those PCEBs containing lower mass secondary stars. The PCEBs in our sample show exactly the  opposite. As our sample is not significantly biased, neither against low-mass secondaries nor long orbital periods, we can exclude the existence of a large currently unidentified population of PCEBs with low-mass (M6-M9) secondaries and long orbital periods (large binary separations, see \citealt{nebot-gomez-moranetal11-1} for details on the orbital period distribution of SDSS PCEBs). This is in clear contrast to the relation proposed by \citet{demarcoetal11-1}.

 A possible explanation for the observed relation between the final orbital period (binary separation) and the companion mass (Fig.\,\ref{fig:M2}) is that systems with massive secondaries have more initial orbital energy available, and therefore a smaller fraction of this energy is enough to unbind the envelope. However, this interpretation is far too simple because the average initial conditions may also depend on the secondary mass. In addition, the relation itself, certainly present in the available data, should be regarded with some suspicion because the currently available sample of PCEBs only covers a relatively narrow range of secondary star masses. More PCEBs with high-mass secondaries are needed to confirm the trend. \\

 \noindent The PCEBs in our sample are consistent with a constant value of the CE efficiency, but each system can be reconstructed using a relatively broad range of values for $\alpha_\mathrm{CE}$ \citep{zorotovicetal10-1}, which implies a wide range of possible initial configurations for each system. We therefore suggest that an even larger and more homogeneous sample is required to evaluate possible dependencies of $\alpha_\mathrm{CE}$ on the binary parameters using reconstruction algorithms.
 
\section{Conclusion}

 The orbital period distribution of PCEBs containing He-core primaries peaks at a significantly shorter orbital period ($\Porb \sim 0.28$\,d) than the one for systems containing C/O-core WDs ($\Porb \sim 0.57$\,d), which is not the result of a continuous increase in orbital period with WD mass. In contrast to recent predictions, we find that the post-CE binary separation increases with the mass of the secondary star. Even though the PCEB sample is still heavily affected by selection effects against detecting PCEBs containing high-mass secondaries, the relations identified here may be able to constrain the energy budget of CE evolution if combined with binary-population-synthesis models that incorporate observational selection effects that are as detailed as possible and that take the different possible combinations of initial conditions and CE efficiencies into account.

 On the observational side, we highly encourage to search for PCEBs containing high-mass secondaries, which is needed to evaluate possible dependencies of $\alpha_\mathrm{CE}$ on the binary parameters based on reconstruction algorithms.

\begin{acknowledgements}
 We acknowledge support from Gemini/Conicyt (grant 32100026, MZ; grant 32090027, NV), FONDECYT (grant 1100782, MRS; grant 3110049, ARM), ESO/Comite Mixto, and the DFG (grant Schw 536/33-1, ADS).
\end{acknowledgements}

\bibliographystyle{aa}
\bibliography{aamnem99,aabib}

\end{document}